\documentclass[doublecol]{epl2} 
\usepackage{epsfig,amsfonts,amsthm}
\usepackage{amsmath,amssymb}
\usepackage{bm}
\usepackage{color}

\newcommand{\be}{\begin{equation}}
\newcommand{\ee}{\end{equation}}
\newcommand{\bea}{\begin{eqnarray}}
\newcommand{\eea}{\end{eqnarray}}

\newcommand{\bk}{{\bf k}}
\newcommand{\bK}{{\bf K}}

\renewcommand{\Re}{\mbox{Re}}

\def\lsim{\mathrel{\rlap{\lower4pt\hbox{\hskip1pt$\sim$}}
    \raise1pt\hbox{$<$}}}         

\def\gsim{\mathrel{\rlap{\lower4pt\hbox{\hskip1pt$\sim$}}
    \raise1pt\hbox{$>$}}}         

\title{Double-slit experiment in momentum space}

\author{I.~P.~Ivanov\inst{1} \and D.~Seipt\inst{2,3} \and A.~Surzhykov\inst{4,5} \and S.~Fritzsche\inst{2,3}}

\institute{                    
  \inst{1} CFTP, Instituto Superior T\'{e}cnico, Universidade de Lisboa,
av.~Rovisco~Pais~1, 1049--001, Lisboa, Portugal\\
  \inst{2} Helmholtz Institut Jena, D--07743 Jena, Germany\\
  \inst{3} Theoretisch-Physikalisches Institut, Friedrich-Schiller-Universit\"{a}t Jena, D--07743 Jena, Germany\\
  \inst{4} Physikalisch--Technische Bundesanstalt, D--38116 Braunschweig, Germany\\
  \inst{5} Technische Universit\"at Braunschweig, D--38106 Braunschweig, Germany
}

\pacs{13.60.Fz}{Elastic and Compton scattering}
\pacs{12.20.-m}{Quantum electrodynamics}
\pacs{41.75.Ht}{Relativistic electron and positron beams}

\abstract{
Young's classic double-slit experiment demonstrates the reality of interference 
when waves and particles travel simultaneously along two different spatial paths.
Here, we propose a double-slit experiment in momentum space, 
realized in the free-space elastic scattering of vortex electrons. 
We show that this process proceeds along two paths in momentum space, 
which are well localized and well separated from each other. 
For such vortex beams, the (plane-wave) amplitudes along the two paths acquire adjustable phase shifts and produce
interference fringes in the final angular distribution.
We argue that this experiment can be realized with the present day technology. 
We show that it gives experimental access to the Coulomb phase,
a quantity which plays an important role in all charged particle scattering but which
usual scattering experiments are insensitive to.
}

\begin{document}
\maketitle

\section{Introduction}
\begin{figure}[!htb]
\centering
\includegraphics[width=0.45\textwidth]{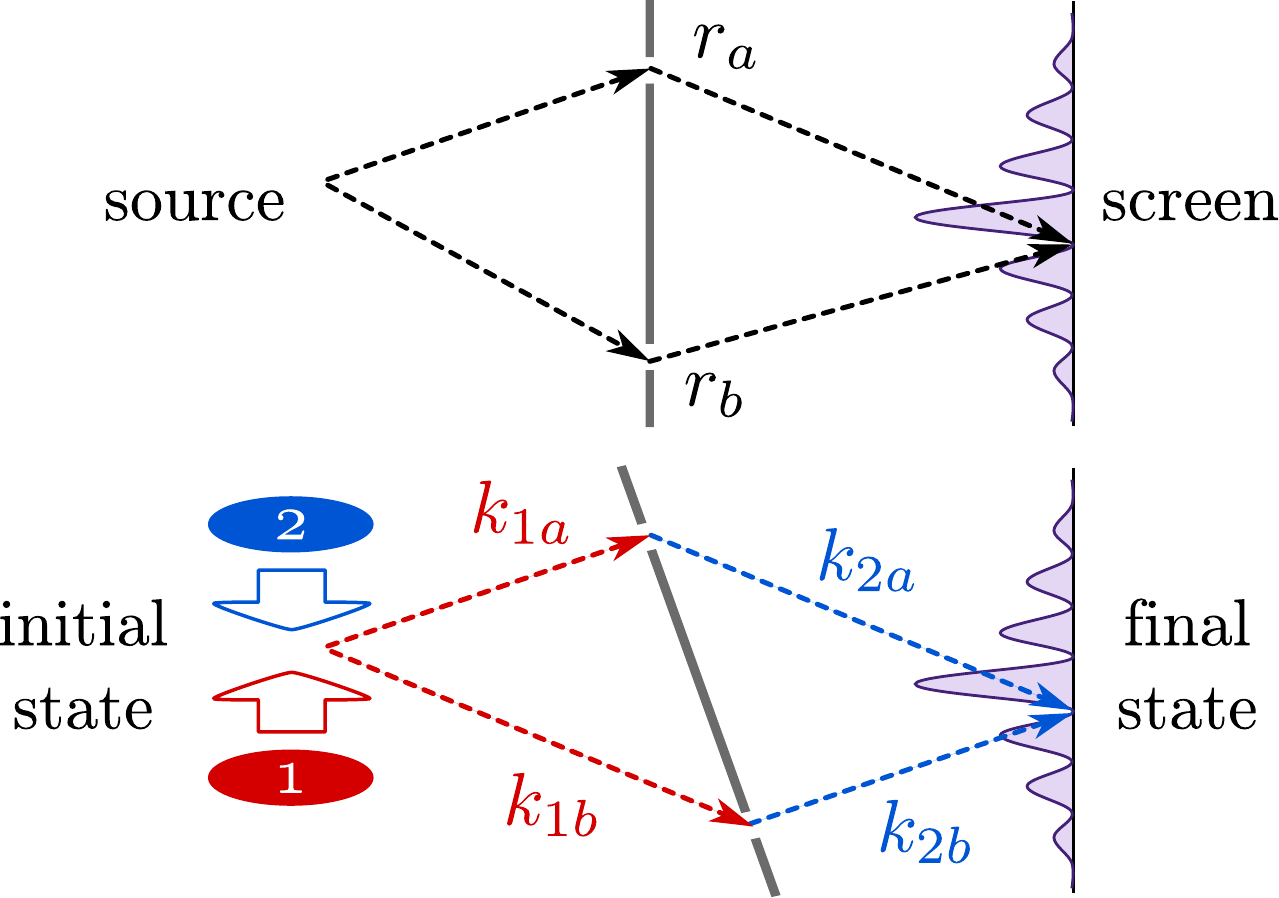}
{\caption{\label{fig-two-slit} Schematic illustration of the Young's experiment in the coordinate space (upper pane)
and of the proposed double-slit experiment in momentum space (lower pane). In the latter case,
we stress that, in the collision of specially prepared wave-packets, only two momentum combinations
lead to a given final state.}}
\end{figure}

Young's seminal double-slit experiment demonstrates the interference of -- classical or
quantum  -- amplitudes if the wave propagate along (two) different spatial paths \cite{optics}. 
A wave emitted from some local source passes through two narrow slits
in a plate, located at coordinates $r_a$ and $r_b$, and interferes with itself on a distant screen, cf.~upper panel of Fig.~\ref{fig-two-slit}.
If the transition amplitudes from the source to a given point on the screen along the 
two paths are $f_a$ and $f_b$, the intensity profile on the screen is 
$|f_a + f_b|^2 = |f_a|^2 + |f_b|^2 + 2\,\Re(f_a f_b^*)$, and the interference term
produces a spatially periodic signal on the screen. 
Such interference fringes have been observed not only for photons but also for electrons, atoms \cite{atoms}, 
and even large organic molecules \cite{large-molecules}.
Moreover, any modification of the physical conditions along either path probed by these particles 
gives rise to a shift in the interference pattern,
which is the basis of numerous interferometric techniques \cite{optics}.

A similar situation occurs frequently 
in collision experiments when an initial state evolves 
into a final state via different intermediate states,
such as different excited states of an atom or different virtual particles in high-energy collisions.
Neutrino oscillations \cite{neutrino} is a hallmark example of such dynamics. A neutrino, which is
produced in a state of definite flavour, propagates to the detection point as a superposition of 
three mass eigenstates. For a fixed neutrino energy, they correspond to different momenta,
and their interference causes spatially oscillating probability 
for changing flavour between production and detection point.
Although one sometimes refers to these oscillations as the momentum-space
analog of the two-slit experiment \cite{lipkin}, 
it is important to keep in mind that one observes here interference 
between amplitudes with the same initial and final state kinematics 
but with different {\em state-space evolution}.
Similar remarks apply to the interference inside atom interferometers \cite{atom}.
A freely falling atom or an atomic cloud is brought into a superposition of excited states,
which owing to their energy difference move with different momenta and 
when recombining produce an interference pattern.

In condensed matter, one encounters examples of interference 
between different momentum-space configurations of the same (quasi)particle along 
the same spatial path. Due to complicated dispersion law,
a definite-energy electron propagating in medium can have two different (quasi)momenta.
These two momentum-paths interfere, leading to a spatially varying electron density profile, 
for example, in the form of spatially modulated tunneling conductance 
in certain superconductors \cite{modulated-conductance}.
In this case, it is the medium that plays the instrumental role as it can absorb the extra momentum 
without destroying the coherence.

In this Letter, we propose a particularly simple momentum-space analog of the Young's double-slit set-up,
which exhibits interference between two amplitudes with identical state-space evolution 
but with different combinations of momenta.
This effect arises in two-particle elastic scattering in free space, 
without any medium to support the evolution,
and uses elementary particles, so that no excitation of internal degrees of freedom is needed.
Such an experiment must involve a scattering process which brings an initial state to the final state 
along two distinct, well-separated paths in the momentum space, as illustrated in the
lower panel of Fig.~\ref{fig-two-slit}.
Each path resembles a plane-wave scattering with its own momentum transfer,
but the final-state kinematics for all particles involved must be identical in order for interference to take place.

We will show that this peculiar set-up is naturally realized in elastic scattering
of electron vortex beams (or twisted electrons), the electron states with helical wave fronts,
which carry non-zero orbital angular momentum projection on the
average propagation direction.
Following the suggestion of \cite{Bliokh:2007}, such electron states were realized experimentally a few years ago
\cite{twisted-electron}, for a recent review see \cite{review}. 
Since the interference occurs in free space, it is free from medium effects and 
can reveal novel features of the fundamental scattering process.
In particular, we will show that it provides direct access the Coulomb phase, a quantity that affects
all charged-particle scattering processes and has been a subject of debates but which
has never been measured experimentally.

\section{The idea}
Consider elastic scattering of two electrons,
each represented by a monochromatic wave-packet 
\be
\psi_i(\vec r) = \int {d^3 k_i}\, a_i(\vec k_i) \exp(i \vec k_i \vec r)\,, 
\label{initial-states}
\ee
with the Fourier amplitude $a_i(\vec k_i)$ being a compact function centered 
at an average momentum $\langle \vec k_i\rangle$.
Scattering of two such electrons 
into a final plane-wave state with momenta $k_1'$ and $k_2'$ is well approximated by the plane-wave 
scattering amplitude ${\cal M}(\langle k_1\rangle, \langle k_2\rangle; k_1', k_2')$,
the cross sections being proportional to its square, $d\sigma \propto |{\cal M}|^2$.
In momentum space, this process is analogous to a {\em single-slit} experiment.

To propose a double-slit experiment in momentum space, we select an axis $z$ and assume that each initial electron
is a monochromatic superposition of plane waves with equal longitudinal momentum $p_{z}$,
equal modulus of the transverse momentum $\varkappa$,
but with different azimuthal angles:
\be
a_i(\vec k_i) \propto \delta(k_{iz}-p_{iz})\delta(|\bk_i| - \varkappa_i) e^{i\ell_i \phi_i}\,,\quad i = 1,2\,.\label{ai}
\ee
Here and below, transverse vectors are denoted with bold letters.
Such states are known as Bessel vortex states, or twisted electrons, \cite{Bliokh:2007,review}.
The parameter $\ell$ can be adjusted experimentally and plays the role of approximately conserved 
orbital angular momentum with respect to the average propagation direction. 
Twisted electrons with energies up to 300 keV are now routinely created 
and are used to gain novel insights into the electron dynamics in external fields,
such as an interplay of Larmor and Gouy rotation in longitudinal magnetic field suggested in \cite{Bliokh:PRX}
and studied experimentally in \cite{Guzzinati:2013,NatureComm}
or the acquisition of phase vortex in the field of an artificial magnetic monopole \cite{monopole}.
More details on technical aspects of generation and manipulation of the vortex electron beams 
can be found in \cite{thesis}.

If the two twisted electrons are brought in collision, they can scatter elastically into a final state
with momenta $k_1'$ and $k_2'$.
Because of momentum conservation, only two plane-wave components with initial momenta $k_{1a}, k_{2a}$ and $k_{1b}, k_{2b}$
lead to this final state.
The total scattering amplitude is then written as a coherent sum of two plane-wave amplitudes: 
\be
f\propto c_a {\cal M}_a + c_b {\cal M}_b\,,\quad {\cal M}_{a,b} = 
{\cal M}(k_{1a,b},k_{2a,b};k_1',k_2')\,.\label{M-generic-ab}
\ee
These two kinematical configurations correspond to different momentum transfers $q = k_1-k_1'$: 
$q_a \not = q_b$. 
The coefficients $c_a$ and $c_b$ in (\ref{M-generic-ab}) depend both on the initial wave packets 
and on the final momenta. In particular, by scanning over the allowed region of $k_1'$ and $k_2'$,
one can change the relative phase between $c_a$ and $c_b$ and observe a periodically varying intensity pattern
in the momentum space. 
The exact position of this pattern in the final-state momentum space 
is sensitive to the phase difference between ${\cal M}_{a}$ and ${\cal M}_{b}$.
This is analogous to the intensity stripes seen on a distant screen in the usual double-slit experiment.

\section{Elastic scattering of Bessel vortex electrons}
We describe the elastic scattering of two Bessel vortex electrons within the fully relativistic 
quantum-electrodynamical framework. Details of this calculations are reported in
\cite{later}; here, we briefly outline the procedure.
Each initial electron is taken as the exact monochromatic 
solution of the Dirac equation with definite energy $E_i$, longitudinal momentum $k_{iz}$, 
helicity $\lambda_i$, and the half-integer total angular momentum $m_i$.
In this work, we use the expressions from \cite{SIFSS-2015}; alternative forms
of these solutions also exist \cite{Bliokh-2011,Karlovets-2012}.
We assume that the two colliding vortex electrons are aligned, that is,
they are defined with respect to the same $z$ axis.
We then select the reference frame in which the longitudinal momenta
are balanced: $k_{2z} = - k_{1z}$, but where the other parameters can still be different from each other: $m_1 \not = m_2$, 
$\varkappa_1 \not = \varkappa_2$, and therefore $E_1 \not = E_2$. 

The two final electrons are plane waves with four-momenta $k^{\prime}_1$ and $k^{\prime}_2$.
Their longitudinal momenta are also balanced, $k_{2z}' = - k_{1z}'$, 
and their energies satisfy $E_1' + E_2' = E_1+E_2$.
The final transverse momenta, however, are {\em not} required to sum up
to zero or to any fixed vector, 
because the initial electrons do not carry definite transverse momenta.
The only kinematical restriction 
is that the total final transverse momentum $\bK$ lies within a ring that is
defined by $\varkappa_1$ and $\varkappa_2$ \cite{ivanov-2011}:
\be
|\varkappa_1 - \varkappa_2| \le |\bK| \le \varkappa_1 + \varkappa_2\,,\quad \bK = \bk_1' + \bk_2'\,.\label{ring}
\ee
For such a scattering of two Bessel beams, the final phase space grows from
the single-particle angular distribution $d\Omega$ or the transverse momentum $d^2\bk_1'$
to the four-dimensional transverse momentum space $d^2\bk_1' d^2\bk_2' = d^2\bk_1' d^2 \bK$.
As a consequence, further information can be extracted from the structures in the final kinematical distributions.

\begin{figure}[h]
\centering
\includegraphics[width=0.49\textwidth]{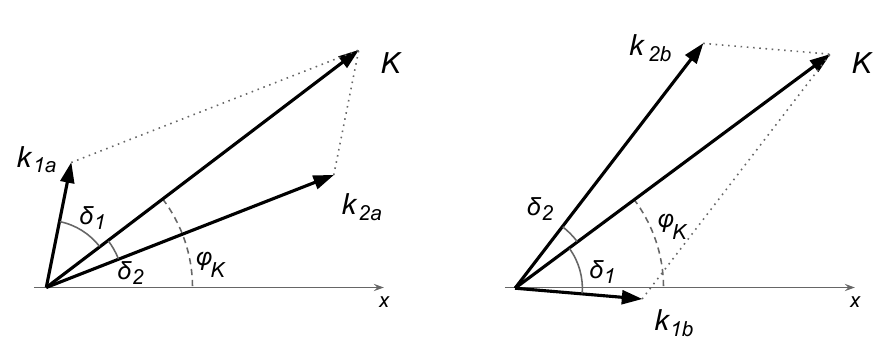}
{\caption{\label{fig-2configurations} The two kinematical configurations in the transverse plane that satisfy
momentum conservation laws in the two Bessel state scattering.}}
\end{figure}

Next, we express the twisted electron scattering matrix element $S_{tw}$ 
as the integral of the plane-wave matrix element $S_{PW}$ over the initial electron wave function:
\be
S_{tw} \propto 
\int d^3 k_1 d^3 k_2\, a_{\varkappa_1 m_1}(\vec k_1)  a_{\varkappa_2 m_2}(\vec k_2) 
S_{PW}\,. \label{Stw0}
\ee
Here
\be
S_{PW} \propto \delta^{(4)}(k_1 + k_2 - k'_1 -k'_2)\cdot
{\cal M}(k_1,k_2;k_1',k_2') \label{SPW}
\ee
and the invariant amplitude ${\cal M}$ is calculated according to the standard Feynman rules \cite{LL}.
For vortex beams, in practice, the calculations of the scattering cross section require special care 
since the regularization and normalization prescriptions,
as well as the definition of the flux differ from the plane-wave case \cite{JS-2011,ivanov-2011,Karlovets-2012,ivanov-2012}. 
In our calculations, all these technical aspects are taken into account, 
but we omit them here for brevity; they are fully presented in \cite{later}.
The integral (\ref{Stw0}) contains equal number of integrations and delta functions,
and it can be worked out exactly. For each $\bK$, it receives contributions from {\em exactly two}
choices of the initial transverse momenta $\bk_{1,2}$ with moduli $\varkappa_{1,2}$ and the azimuthal angles
$\phi_{1a} = \phi_K + \delta_1$, $\phi_{2a} = \phi_K - \delta_2$ and
$\phi_{1b} = \phi_K - \delta_1$, $\phi_{2b} = \phi_K + \delta_2$,
where
\be
\delta_{1,2} = \arccos\left({\varkappa_{1,2}^2 + \bK^2 - \varkappa_{2,1}^2 \over 2\varkappa_{1,2} |\bK|}\right)\label{deltas}
\ee
are the internal angles of the triangle with sides $\varkappa_1$, $\varkappa_2$, $|\bK|$.
The area of this triangle is denoted by $\Delta = |\bK|\varkappa_1 \sin\delta_1/2$.
These two configurations are shown in Fig.~\ref{fig-2configurations}.

Squaring the matrix element, regularizing it, and dividing by the flux, we obtain the following expression
for the twisted electron cross section:
\be
d\sigma_{tw} \propto |{\cal J}|^2 d^2 \bk'_1 d^2 \bk'_2 = |{\cal J}|^2 d^2 \bk'_1 d^2 \bK\,,\label{dsigma-tw}
\ee
where
\bea
{\cal J} &=& \int d^2 \bk_1 d^2 \bk_2 \delta^{(2)}(\bk_1+\bk_2 - \bK) \nonumber\\
&&\times\  \delta(|\bk_1|-\varkappa_1) \delta(|\bk_2|-\varkappa_2)
\nonumber\\
&&\times\  e^{im_1\phi_1 - im_2\phi_2}{\cal M}(k_1,k_2;k_1',k_2') \label{J} \\
&=& e^{i(m_1 - m_2)\phi_K}{\varkappa_1 \varkappa_2 \over 2\Delta}\times \nonumber\\
&\times&\left[{\cal M}_{a}\, e^{i (m_1 \delta_1 + m_2 \delta_2)} 
+ {\cal M}_{b}\, e^{-i (m_1 \delta_1 + m_2 \delta_2)}\right].\label{J2}
\eea
Here, ${\cal M}_{a}$ and ${\cal M}_{b}$ are the two plane-wave scattering amplitudes as in Eq.~(\ref{M-generic-ab}),
which are calculated for the two distinct initial momentum configurations shown in Fig.~\ref{fig-2configurations}. 
They correspond to the same relativistic $s$-invariant but two distinct $t$-invariants
$t_{a,b} = (k_{1a,b} - k_1')^2 = (k_{2a,b} - k_2')^2$, whose difference is
\be
t_a - t_b = 2\bk_1' (\bk_{1a}-\bk_{1b}) = 4 |\bk_1'| \varkappa_1 \sin\delta_1 \sin(\phi_1' - \phi_K)\,.\label{tatb}
\ee

\section{Interference fringes}
To get qualitative understanding, consider (\ref{dsigma-tw}) in
the ultrarelativistic small scattering angle approximation: $|\bk_i|,\, |\bk_i'| \ll k_z$.
Unlike the plane-wave scattering, where the total transverse momentum fulfills the condition $\bK=\bk'_1+\bk'_2 = 0$, 
we have here an extra dimension to look at: the $\bK$-distribution {\em at fixed $\bk'_1$}.
This distribution fills the ring defined by (\ref{ring}) and display interference fringes.
For high electron energies and small scattering angles, the helicity amplitude can be approximated by 
\be
{\cal M} = 8 \pi \alpha_{em} {s \over t} e^{-i\lambda_1 (\phi_1-\phi_1')} e^{i\lambda_2 (\phi_2-\phi_2')}
\delta_{\lambda_1 \lambda_1'} \delta_{\lambda_2 \lambda_2'}\,.\label{Born-UR}
\ee
Substituting it into (\ref{J2}), we obtain, for the unpolarized case,
\bea
&&{1 \over 4}\sum_{\lambda_i} |{\cal J}|^2 = 64 \pi^2 \alpha^2_{em} s^2
{\varkappa_1^2\varkappa_2^2 \over 4\Delta^2}\label{Jsquared}\\
&\times&\left[{1 \over t_a^2} + {1 \over t_b^2} 
+ {2 \over t_a t_b}\cos(2m_1\delta_1+2m_2\delta_2)\cos\delta_1\cos\delta_2\right].\nonumber
\eea
By detecting electrons with different $\bk_2'$ at fixed $|\bk_1'|$,
we can scan the ring (\ref{ring}). As seen from (\ref{deltas}), the angles $\delta_i$ change,
and the last term in (\ref{Jsquared}) oscillates, producing concentric ring structure.
These are the characteristic interference fringes of Young's double-slit experiment but now in
the {\em momentum space}.

Notice that as we move from the outer to the inner boundary of the ring during this scan, 
the two kinematical configurations shown in Fig.~\ref{fig-2configurations} first separate apart
in momentum space and then move towards each other to finally merge at the boundary.
In that aspect, the momentum-space two-slit experiment differs from the usual 
set up with two fixed slits in a plate.

\begin{figure}[h]
\centering
\includegraphics[width=0.45\textwidth]{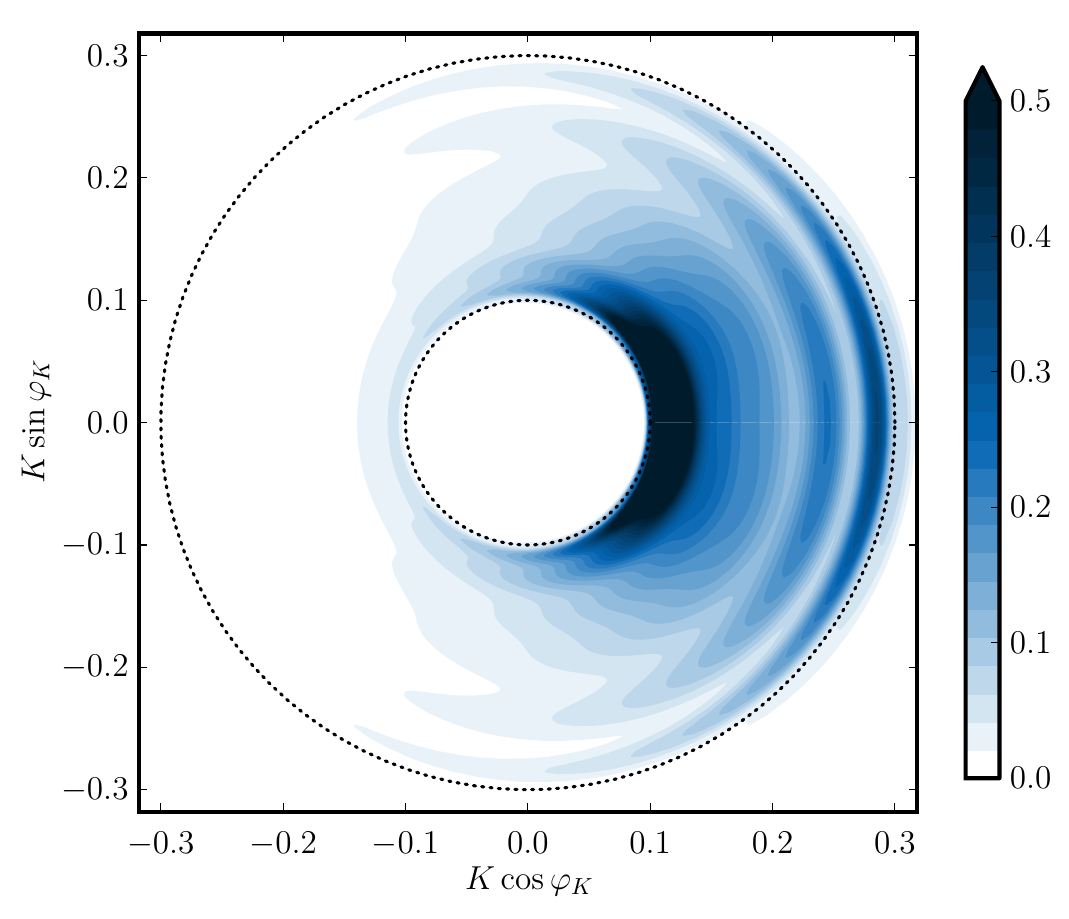}
\includegraphics[width=0.45\textwidth]{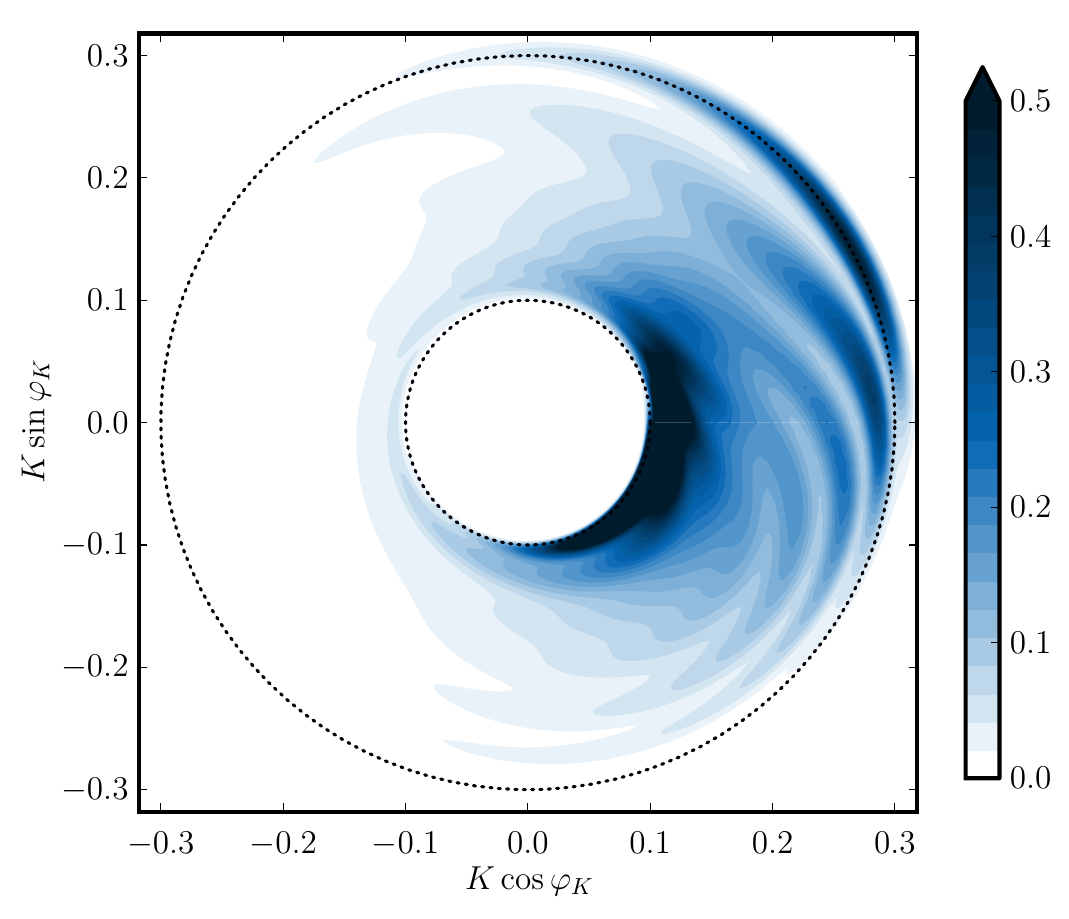}
{\caption{\label{fig-K} Differential cross section, in arbitrary units,
as a function of $\bK$ for fixed $\bk_1'$ for a choice of parameters (see text).
The upper plot is calculated from the purely real Born-level amplitude, while the lower plot
takes into account the Coulomb phase (\ref{zeta}),
with $\alpha_{em}$ artificially set to 10 to enhance the visibility
of the up-down asymmetry.}}
\end{figure}

Before verifying the above analysis with numerical results,
we mention that pure Bessel beams are not normalizable
and, formally, lead to divergent expressions.
However, realistic vortex beams produced in experiment are free from these difficulties.
To model them, we follow \cite{IS-2011,ivanov-2012} and replace the singular Fourier components $a(\vec k)$
as in (\ref{ai}) with their Gaussian-distributed Fourier components
$\tilde a(\vec k) = \int d\varkappa f(\varkappa)a_\varkappa(\vec k)$, where 
$f(\varkappa) \propto \exp\left[-(\varkappa-\bar\varkappa)^2/2\sigma^2\right]$.
This leads to the corresponding averaging over a region of $k_{1z}$ and $k_{2z}$,
so that we get a narrow distribution over final $K_z$ peaked at $K_z=0$, over which we further integrate
the cross section.

The upper pane of Fig.~\ref{fig-K} shows the numerically obtained cross sections 
with the following parameters:
$E_1 = 2.1$ MeV, $\bar\varkappa_1 = 200$ keV, $\bar\varkappa_2 = 100$ keV, $\sigma_i = \bar\varkappa_i /20$,
$|\bk_1'| = 500$ keV, $m_1 = 1/2$, $m_2 = 13/2$.
In this plot, $\bk_1'$ is directed to the right.
One sees the interference fringes as well as the strong left-right asymmetry, 
originating from the $t$-dependence of (\ref{Jsquared}).
The plot is symmetric with regard to the horizontal line because 
the cross section contains no term proportional to $\sin(\phi'_1-\phi_K)$. 

The visibility of interference fringes depends 
on the parameters of twisted electrons. 
The relative width of the ring is determined by the ratio $\varkappa_1/\varkappa_2$,
while the number of fringes is sensitive to both $\varkappa$'s and $m$'s.
The parameter $\sigma$ of the Gaussian function, whose inverse is essentially the transverse coherence length, 
works against narrow fringes, smearing the interference pattern with too many fringes.
Our numerical study shows that in order to produce a strong fringe contrast such as shown in Fig.~\ref{fig-K} 
with a reasonably small $\sigma$,
one needs to choose comparable $\varkappa_1, \varkappa_2$ and the values of $m_i$ should be
of several units \cite{later}. We do not expect that large $m_i$, which are feasible experimentally, 
will improve the contrast.

\section{Accessing the Coulomb phase}
The two-slit interferometry in momentum space as proposed here provides direct access to a quantity
which cannot be measured in the usual plane-wave scattering:
the phase of the (complex) scattering amplitude,
or more accurately, its dependence on the momentum transfer squared $t$.
For elastic scattering of charged particles, this phase is known as the Coulomb phase.
If we write the amplitude as ${\cal M} = |{\cal M}| e^{i\zeta}$, 
one obtains for ultrarelativistic same-sign charged particles
\be
\zeta = \zeta_0 + \alpha_{em} \ln (1/|t|)\,,\label{zeta}
\ee
where $\zeta_0$ is the unobservable overall phase which depends on the infrared regularization.

The overall phase of any scattering process cannot be measured experimentally 
because it drops out of the plane-wave cross section:
$d\sigma \propto |{\cal M}|^2$.
If the scattering is entirely due to electromagnetic interactions, 
this phase is equal to the Coulomb phase and is unobservable.
However, the Coulomb phase does affect the elastic scattering of charged hadrons.
Its amplitude receives contributions from the strong and electromagnetic interactions, 
${\cal M} = {\cal M}_s + {\cal M}_{em}$, each coming with its own phase.
Knowing the Coulomb phase as accurately as possible is needed to 
probe the large unknown strong phase via the interference between the strong and electromagnetic amplitudes.
The problem is complicated by the fact that the two contributions
become interrelated at higher orders of the perturbation series.
The strong amplitude receives multi-photon exchange corrections,
and the electromagnetic amplitude involves intermediate excited hadronic states.
These issues sparked debates back in 1960's \cite{Bethe-1958,Rolnick-1966,rix-thaler-1966,islam-1967,solovyev-1966,west-yennie-1968},
and are still discussed today, in the context of elastic $pp/p\bar p$ and deep-inelastic scattering \cite{cahn-1982,selyugin-1999,kopeliovich-2001,prokudin-2002,dremin-2013}.

The two-slit interferometry in momentum space may lead to the first ever measurement of the $t$-dependence
of the Coulomb phase. This occurs since the interference term in (\ref{M-generic-ab})
is sensitive to the phase difference,
\be
d\sigma_{int} \propto 2|{\cal M}_a||{\cal M}_b|\Re\left[c_a c_b^* e^{i(\zeta_a - \zeta_b)}\right]\,,
\ee 
where $\zeta_a - \zeta_b \approx \alpha_{em} \ln(t_b/t_a)$.
The extra phase difference between ${\cal M}_a$ and  ${\cal M}_b$ distorts the interference pattern.
The cross section acquires a new term proportional 
to $t_a-t_b \propto \sin(\phi_1' - \phi_K)$ and leads to the up-down asymmetric
$\bK$-distribution, which is entirely due to the $t$-dependent Coulomb phase.
The lower pane of Fig.~\ref{fig-K} shows this asymmetry for the same parameter set as before.
Note that here, for the sake of illustration, the effect is greatly exaggerated by setting $\alpha_{em} = 10$.
For physical $\alpha_{em}$, the effect is propostionally smaller, 
but it can be extracted from the experimentally measured cross section
via asymmetry $A_{\perp} = \int d\sigma_{tw} \sin(\phi_1' - \phi_K)/ \int d\sigma_{tw}$.
Our numerical calculations give $A_\perp = {\cal O}(10^{-4}\div10^{-3})$, whose smallness is mostly driven by the small $\alpha_{em}$. 
The exact value of $A_\perp$ strongly depends on the details of the initial state;
by adjusting them, one can further optimize the sentisivity of this measurement to the Coulomb phase.

\section{Which-way experiment in the momentum space}
The ``which way experiment'' is a variation of the classical double-slit experiment, in which  
a device is placed next to one slit in order to detect whether the particle actually passes through it. 
Using such a device lets interference disappear
or, for non-ideal detection, degrade, displaying the laboratory proof of the reality 
of quantum-mechanical complementarity \cite{which-way}.
The proposed two-slit experiment in momentum space can also be modified 
in order to establish a corresponding which-way experiment.
The idea is to consider {\em inelastic} scattering, $ee\to ee\gamma$, 
in which a sufficiently energetic bremsstrahlung photon is 
detected in coincidence with the scattered electrons. 
Since the two interfering plane-wave configurations have different initial momenta,
say $k_{1a}$ and $k_{1b}$, detection of the bremsstrahlung photon close to the direction of $k_{1a}$
gives preference to this ``slit'' in the momentum space. Selecting different photons,
one can change the efficiency of the ``which-way detection'', 
and the interference fringe contrast must vary accordingly.

\section{Discussion and conclusions}
The proposed experiment can be realized with present day beams and detectors. 
Vortex electrons with energies up to 300 keV and focused to angstrom-size focal spots
are now routinely produced and manipulated in electron microscopes \cite{thesis}.
Scattering of two vortex electron beams has not yet been studied experimentally,
but it can be readily done once the instrumentation is modified for this purpose.
Building such a device with two independently operational electron microscopes
producing counterpropagating vortex electron beams, in which each microscope
is protected against the opposite beam and offers enough space
to install the scattered electron detectors, represents the main technical challenge 
on the way to experimental realization of this experiment.
Although our calculations were done for the ideal situation of perfect alignment of the two counterpropagating beams,
numerical estimates show that the interference picture remains even under slight misalignment
provided the shift between the two axes is kept within the focal spot radius 
and the tilt is less than the cone opening angle $\varkappa/|k_z|$ \cite{later}. 
The present state-of-the-art manipulation with vortex electron beams
offers the required level of control, so that 
stable alignment of the two beams within a common angstrom-scale focal spot seems feasible.
With the kinematical parameters used for illustration, we estimate the scattering probability
for each $ee$ crossing of about $P \sim \sigma_{tw}/S_{\rm focal} \sim 10^{-6}$.
With the beam currents of 1 nA, one can set up billions of $ee$ collision attempts
per second resulting in hundreds of detectable scattering events per second.
A few hours of observation time will produce a million-event statistics. 

In order to construct plots such as Fig.~\ref{fig-K}, the detectors must
detect scattered electron pairs in coincidence and with sufficient angular resolution,
while the electron energy need not to be measured explicitly. 
From these coincidence measurements, one then slices
the full sample of detected $\bk_1'$ and $\bk_2'$ pairs into
subsamples with given $|\bk_1'|$, reconstructs $\bK$,
and plots the events in the $\bK$-space with respect to the orientation of $\bk_1'$.
A million-event statistics should be enough 
to detect the interference fringes inside the annular $\bK$-region.
In order to detect a non-zero asymmetry $A_\perp$ and probe the Coulomb phase,
these structures need to be measured with even higher accuracy, 
which seems challenging at present.

To conclude, we proposed the momentum-space analogue of the classical Young's double-slit experiment,
which occurs in free-space collision of specially prepared states.
When two Bessel vortex electrons scatter elastically, the process is dominated by
two plane-wave scattering configurations with different momentum transfers.
Just like two slits in the usual Young's interference experiment,
these two configurations corresponds to two paths which are well localized and well separated in the momentum space.
The two paths acquire adjustable phase difference, their amplitudes sum up coherently
and lead to interference fringes in the final state angular distribution.
We showed, in particular, that this interference allows for the first ever experimental investigation
of the Coulomb phase.
We also proposed the which-way modification of this experiment, with a bremsstrahlung
photon playing the role of a detecting device with imperfect efficiency.
We argued that the proposed momentum-space double-slit experiment can be realized with the present day technology. 

\bigskip

The work of I.P.I. was supported by the Portuguese
\textit{Fun\-da\-\c{c}\~{a}o para a Ci\^{e}ncia e a Tecnologia} (FCT)
through the FCT Investigator contract IF/00989/2014/CP1214/CT0004
under the IF2014 Programme, as well as
under contracts UID/FIS/00777/2013 and CERN/FIS-NUC/0010/2015,
which are partially funded through POCTI, COMPETE, QREN, and the EU.
I.P.I. is also thankful to Helmholtz Institut Jena for hospitality during his stay
as a Visiting Professor funded
by the ExtreMe Matter Institute EMMI, 
GSI Helmholtzzentrum f\"{u}r Schwerionenforschung, Darmstadt. 
S.F. acknowledges support by the QUTIF priority programme of the DFG (FR 1251/17-1).

\end{document}